\documentclass{elsart}

\usepackage{harvard}

 \usepackage{graphics}
\usepackage{graphicx}
\usepackage{epsfig}

\usepackage{amssymb}

\def\url#1{{\ttfamily\def\/{/\discretionary{}{}{}}#1}}

\def\sasz{{\tt SASZ~{}}}
\def\aap{A\&A}
\def\apj{ApJ}
\def\apjs{ApJS}
\def\prd{PRD}
\def\apjl{ApJL}
\def\mnras{MNRAS}

\begin{document}

\begin{frontmatter}
\title{Extraction of cluster parameters from Sunyaev-Zeldovich effect observations with simulated annealing optimization}
\author{Steen H. Hansen}
\address{University of Zurich, Winterthurerstrasse 190,
8057 Zurich, Switzerland}
% use the thanksref command within \title, \author or \address for footnotes:
% \title{\thanksref{label1}}
% \thanks[label1]{}
% \author{\thanksref{label2}}
% \thanks[label2]{}
% \address{\thanksref{label3}}
% \thanks[label3]{}
% including your email address
% \address{\thanksref{email}}
% \thanks[email]{E-mail: }

\begin{abstract}
We present a user-friendly tool for the analysis of data
from Sunyaev-Zeldovich effect observations. The tool is based on the
stochastic method of simulated annealing, and allows the extraction of
the central values and error-bars of the 3 SZ parameters,
Comptonization parameter, $y$, peculiar velocity, $v_p$, and electron
temperature, $T_e$. The f77-code \sasz will allow any number of
observing frequencies and spectral band shapes.
As an example we consider the SZ parameters for the COMA cluster.
%, and is easily extended to include other physical parameters.
\end{abstract}

%\begin{keyword}
% keywords here, in the form keyword \sep keyword
% PACS code here, in the form \PACS code \sep code
%Intergalactic medium -- Galaxies: cluster: general
%\PACS 95.75.Pq \sep 98.65.Cw
%\end{keyword}
\end{frontmatter}

% main text

\section{Introduction\label{sec:intro}}

Galaxy clusters typically have temperatures of the order keV, $T_e =
1-15$ keV, and a CMB photon which traverses the cluster and
happens to Compton scatter off a hot electron will therefore get increased
momentum. This up-scattering of CMB photons, which results in a small
change in the intensity of the cosmic microwave background, is known
as the Sunyaev-Zeldovich (SZ) effect, and was predicted just over 30 years
ago~\cite{sz72}.  The first radiometric observations came few years
later~\cite{gull76,lake77}, and while recent years have seen an
impressive improvement in observational techniques and
sensitivity~\cite{laroque02,coma}, then the near future observations
will see another boost in sensitivity by orders of magnitude. These
include dedicated multi-frequency SZ observations like
ACT~\footnote{{\tt http://www.hep.upenn.edu/$\sim$angelica/act/act.html}}
and SPT~\footnote{{\tt http://astro.uchicago.edu/spt/}}. 
The SZ effect will thus soon provide us with an
independent description of cluster properties, such as evolution and
radial profiles. For recent excellent reviews see~\cite{birk99,carl02}.

The SZ effect is traditionally separated 
into two components according
to the origin of the scattering electrons
\begin{eqnarray}
\frac{\Delta I(x)}{I_0} &=& \Delta I_{{\rm thermal}} \, (x, y, T_e)
+ \Delta I_{{\rm kinetic}} \, (x, \tau, v_p,T_e)
\label{eq:deltai}
\\ 
&=& y \, \left( g(x)+\delta_T(x,T_e) \right) 
-  \beta \tau \, \left(h(x) + \tilde \delta_{kin} (x,T_e)  \right) 
\,, \nonumber
\end{eqnarray}
with $x = h \nu/kT_{cmb}$ and $I_0 = 2 (kT_{cmb})^3/(hc)^2$
where $T_{cmb} = 2.725$ K. 
The first term on the rhs of eq.~(\ref{eq:deltai})
is the thermal distortion with the non-relativistic spectral shape
\begin{equation}
g(x) = \frac{x^4 \,e^x}{(e^x -1)^2} \left( x\,\frac{e^x + 1}{e^x -1} -
4 \right)\,,
\end{equation}
and the magnitude is given by the Comptonization parameter 
\begin{equation}
\label{eq:ygas}
y = \frac{\sigma_{\rm T}}{m_e \, c^2}\, \int\!\! dl\, n_{e}
\,kT_e \, ,
\end{equation}
where $m_e$ and $n_e$ are masses and number density of the electrons,
and $\sigma_T$ is the Thomson cross section.

For non-relativistic electrons one has $\delta_T(x,T_e) = 0$, but
for hot clusters the relativistic electrons will slightly modify
the thermal SZ effect~\cite{1979ApJ...232..348W}. 
These corrections are easily 
calculated~\cite{1995ApJ...445...33R,itohnozawa03,2000A&A...360..417E,2001ApJ...554...74D}, and can be used to measure the
cluster temperature purely from SZ 
observations~\cite{2002ApJ...573L..69H}.
For the implementation below we will use an extension of the method
developed in~\citeasnoun{2003JCAP...05..007A}, using a fit to
the spectral shape of $\delta_T(x,T_e)$, which 
everywhere in the range $20-900$ GHz and $T_e < 24$ keV
is very accurate, $| \delta^{fit}_T - \delta_T | / ( | \delta_T | 
+ |g|) < 0.005$. In the range $24 \, {\rm keV} < T_e < 100$ keV the
accuracy is slightly lower.

The kinetic distortions have the spectral shape
\begin{equation}
h(x) = \frac{x^4 \,e^x}{(e^x -1)^2} \, ,
\end{equation}
and the magnitude depends on $\beta = v_p/c$, the average line-of-sight streaming
velocity of the thermal gas (positive
if the gas is approaching the observer), and the Thomson optical
depth
\begin{equation}
\label{eq:barbeta}
\tau = \sigma_{\rm T_e} \int\!\! dl\, n_{e} \, .
\end{equation}
Thus, when the intra-cluster gas can be assumed isothermal one
has $y = \tau kT_e/(mc^2)$.
For large electron temperatures 
there are also small corrections to the kinematic effect, 
$\tilde \delta _{kin}(x,T_e) 
\neq 0$~\cite{1998ApJ...508....1S,1998ApJ...508...17N}, an effect
which is negligible with present day sensitivity.

Given the different spectral signatures of $g(x), h(x)$ and $\delta_T
(x,T_e)$, it is straight forward to separate the physical variables
$y,v_p$ and $T_e$ from sensitive multi-frequency observation.
However, due to the complexity of the spectral shapes, in particular of
$\delta_T(x,T_e)$, the parameter space spanned by $y,v_p$ and $T_e$ may
be non-trivial with multiple local minima in $\chi ^2$. 
We therefore present a stochastic analysis tool \sasz
based on simulated annealing, which allows a safe and fast parameter
extraction even for such a complex parameter space.

It is worth noting that whereas we here choose to use the set of
cluster variables, $(y, T_e, v_p)$, which are easily understood
physically, then the analysis could be simplified significantly by
introducing a set of {\it normal} parameters whose likelihood function
is well-approximated by a normal
distribution~\cite{2002PhRvD..66f3007K,chu02}. This set of normal
parameters could e.g.\ be $(y, T_e, K=\tau v_p)$, which directly enter
in eq.~(\ref{eq:deltai}). Another set could be $(y, T_e, \tilde K=v_p
T_e^{-0.85})$, where $\tilde K$ enters because the cross-over frequency,
$\nu _0$,
is easily determined observationally (due to a fast variation of
$\Delta I(x)$) and the fact that this
cross-over frequency to a good approximation depends only on $T_e$ and
$\tilde K$
\begin{equation}
\nu _ 0 = 217.4  \, (1 + 0.0114 \, T_5) + 12 \, T_5 ^{-0.85} \,
v_{500} \, {\rm GHz}\, ,
\end{equation}
using $T_5 = T_e /(5 {\rm keV})$ and $v_{500} = v_p/(500 {\rm km/sec})$.

\section{Simulated annealing}

Let us now discuss the stochastic method used 
in \sasz.
The idea behind the technique of simulated annealing is as follows. If
a thermodynamic system is cooled down sufficiently slowly then
thermodynamic equilibrium will be maintained during the cooling phase.
When the temperature approaches zero, $T_A \rightarrow 0$, then the
lowest energy state of the system, 
$E_{min}$,  will be reached~\cite{1983Sci...220..671K}.

One can use this idea to search for the minimum in the space of
allowed parameters, in which case the energy is replaced with
$E = \chi ^2$. The method is related to Monte Carlo methods,
because one is basically jumping randomly around in parameter space,
and if a given new point has lower energy (that is lower 
$\chi_i ^2$ in our case)
than the previous point ($\chi ^2 _{i-1}$),
then this point is accepted.  To ensure that a system is not trapped
in a local minimum there is a certain probability of keeping the new
point even if its energy may be larger than the previous 
one~\cite{metropolis53}
\begin{displaymath}
P_{\rm accept} = \left\{ 
\begin{array}{ll} 1 & \, \, \, \, \, \textrm{if} \, \, \, \, \, 
E_i \leq E_{i-1} \\
e^{-\left( E_i - E_{i-1} \right) /T_A} & \, \, \, \, \, 
\textrm{if} \, \, \, \, \,  E_i > E_{i-1}
\end{array} \right.
\end{displaymath}
and this probability, $P_{\rm accept}$, 
will then be compared to a random number
between 0 and 1, to decide if the point will be kept or not.
The temperature of the system, $T_A$, is then slowly lowered to ensure
that the global minimum $\chi ^2$ is reached. We emphasize that the
simulated annealing temperature, $T_A$, is completely unrelated to
the electron temperature, $T_e$, of the galaxy cluster.

In reality the jumping in parameter space is not completely
random, instead the new points are drawn according to
\begin{equation} 
x_i = x_{i-1} + A_\beta \, \sqrt{\frac{T_A}{T_0}} \, ran
\label{eq:abeta}
\end{equation}
where $ran$ is a random number between $-1/2$ and $1/2$. The
size of the jump is such that initially all of parameter
space is easily sampled, $A_\beta \approx \left( x_{\rm max} -x_{\rm min} 
\right)$, whereas for a cooled system only very small jumps are allowed
due to the $\sqrt{T_A}$ in eq.~(\ref{eq:abeta}).
For flat directions in parameter space (such as $T_e$ and  $v_p$)
the coefficient in $A_\beta$ is a factor 10 larger
than for the dominating Comptonization parameter $y$.
In the first version of \sasz 
$x$ is a 3 dimensional 
vector, $\vec{x} = (y, T_e, v_p)$,
as energy we are using $E = \chi ^2$, and we allow approximately
1000 random jumps at each temperature step.
The cooling scheme is adaptive according to how many point
are accepted or rejected, but can trivially be fixed to an
exponential cooling scheme, $T_A(j) = c \, T_A(j-1)$, where $c<1$.
We use $T_0=1$ and $T_{\rm final} = 10^{-12}$, and the parameter space
allowed is presented in Table 1.

\hspace{0.5cm}

\begin{tabular}{cccc}
\hline \hline
Parameter & Allowed range && Units\\ \hline
$log(y)$ & $-7$  &  $-2$ & \\
$T_e$ & 0 & 100 & keV \\
$v_p$ & $-10^4$ &  $10^4$ &km/sec \\ 
\hline \hline 
\end{tabular}
\hspace{0.5 cm}

We note that the technique of simulated annealing, which is often used
for problems like the travelling salesman problem, has previously been
used is cosmology, e.g.\ for parameter extraction from cosmic
microwave background radiation
data~\cite{1995PhRvD..52.4307K,2000PhRvD..61b3002H}.

\subsection{How to use \sasz in practice?}
The user has a set of observing frequencies (and possibly frequency
bands) with corresponding SZ observations.
The user specifies if some of the parameters have been observed 
with other methods; the temperature can be known to be e.g.\
$T_e = 12 \pm 1$ keV, or the peculiar velocity can be known e.g.\
$v_p = 300 \pm 500$ km/sec. The Fortran77
code \sasz will then stochastically analyse the parameter space (as
described above) and presents the central value and $1\sigma$ error-bars for
the cluster parameters $(y, v_p, T_e)$. The method is much faster than
e.g.\ the method of automatic refining grids~\cite{2003JCAP...05..007A}.
A user guide with examples can be downloaded
together with the 
code~\footnote{{\tt  http://krone.physik.unizh.ch/\~{}hansen/sz/}}. This user
guide also explains how to implement measured frequency band if
available.
%, and how to include other physical parameters like 
%contamination from radio
%point sources or dust.
%~\cite{hansen.aghanim}, 
%or even grey dust~\cite{ensslin.hansen}.

\subsection{Error-bars}
There are two possible steps for determination of error-bars.
The first is a very fast  Gaussian approximation,
and the second is a more accurate method which simultaneously
provides data for nice figures.

The Gaussian estimate of the error-bars comes for free.  While
searching for the global minimum \sasz will remember some intermediate
points, $\vec x_i$, and their $\chi ^2_i$.  For each of the parameters,
e.g. $T_e$, one can consider the 2-dimensional figure $(T_e, \Delta
\chi^2)$, where all points lie within a quadratic curve. One can fit a
quadratic curve through the best fit point and any of the other
points. This method gives a fairly good estimate of the error-bar by
performing such fit for all the points and then choosing the largest
value.  If upper and lower error-bars are different from each other,
then the bigger is chosen for simplicity.  The code selects a certain
number of point for the fit, and presents automatically the error-bars
for the 3 SZ parameters.

The second method is to make a grid in the parameter we are
interested in, 
and then minimize over the other parameters for each point on
that grid. In this way one can 
get different upper and lower error-bars more accurately. This 
method is, however, slower because one will have to run another
optimization for each chosen point in the grid. 
\sasz automatically volunteers to
perform such a calculation in the $2\sigma$ range found with the 
Gaussian method or within a user-defined range.

\section{An example: COMA}

As an example of the use of \sasz we consider the nearby cluster
COMA at redshift, $z=0.0231 \pm 0.0017$. This cluster has been 
measured at 6 different frequencies, 32 GHz~\cite{herbig95}, 
143, 214 and 272 GHz~\cite{coma}, and at 61 and 94 
GHz~\cite{2003ApJS..148...97B}. For a thorough discussion of consistency
and a combined analysis see~\citeasnoun{Battistelli:2003eh}, where
the results are combined in their Table~1.

\begin{figure}[tbh]
\begin{center}
\includegraphics[width=6.5cm,angle=0]{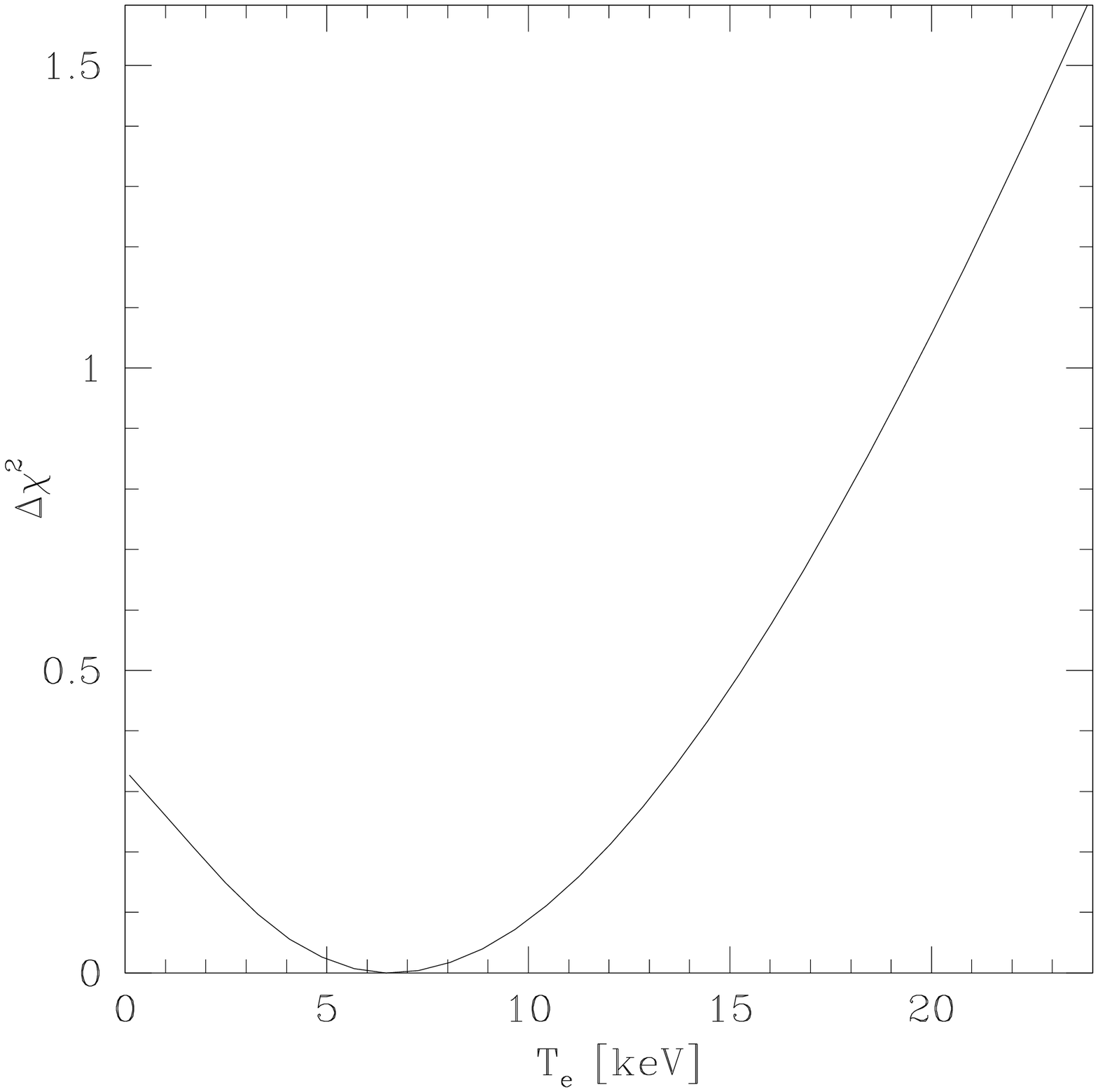}
\includegraphics[width=6.5cm,angle=0]{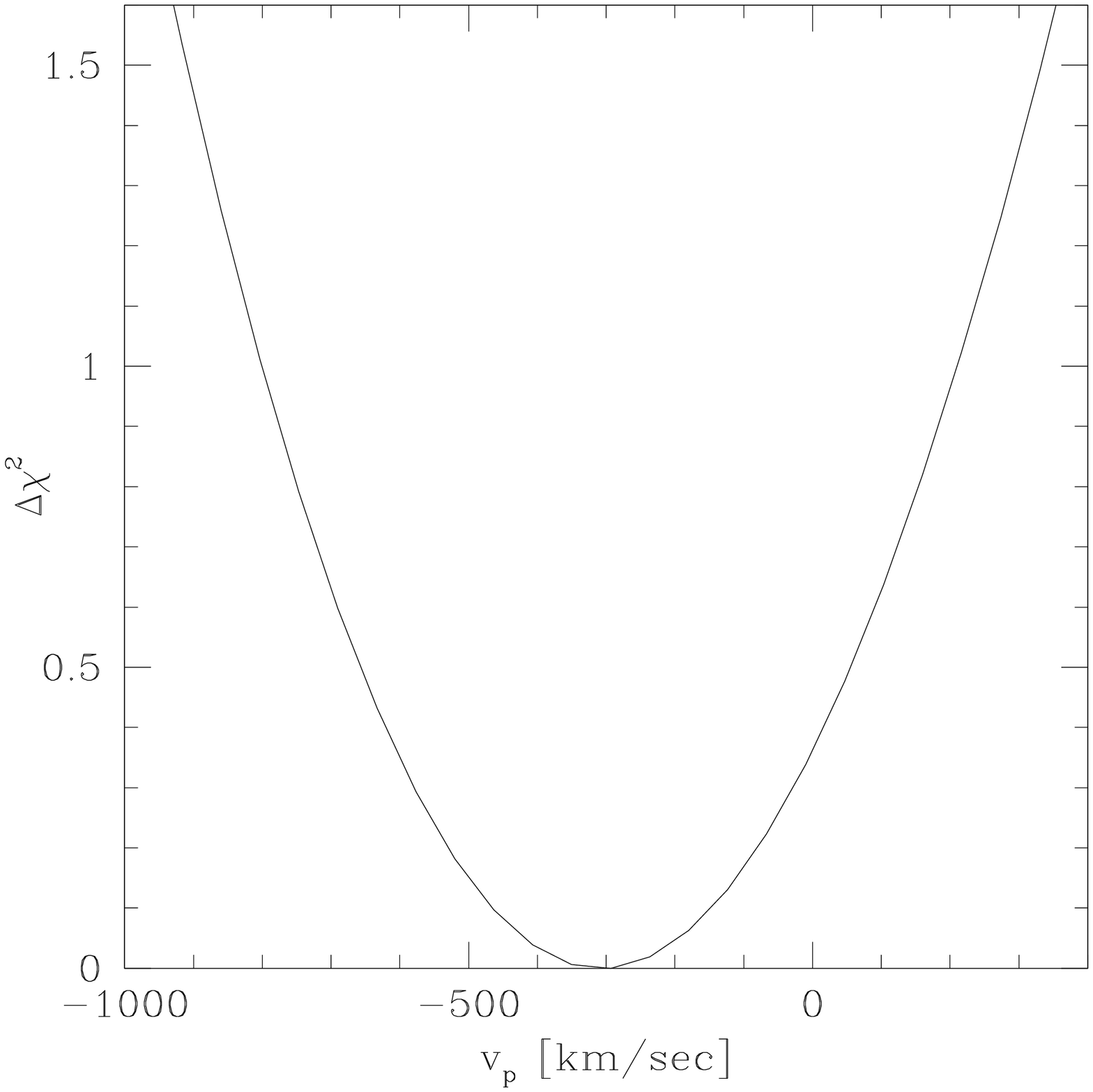}
\end{center}
\caption{Using \sasz to determine the electron temperature or
peculiar velocity for COMA. 
Figure 1a shows 
$\Delta \chi^2$ as a function of the electron temperature, $T_e$.
Other parameters have been maximized over. The $1\sigma$ error-bar on
$T_e$ (corresponding to $\Delta \chi^2 = 1$) gives $T_e < 20$ keV.
Figure 1b shows $\Delta \chi^2$ as a function of the peculiar velocity, $v_p$.
Other parameters have been maximized over. The $1\sigma$ error-bar on
$v_p$ (corresponding to $\Delta \chi^2 = 1$) gives $v_p = -300 \pm 500$
km/sec.}
\label{fig:coma.temp}
\end{figure}

Let us first consider the effect of bandwidth on the determination of $y$. 
We look at the
case where the temperature is measured, 
$T_e = 8.25 \pm 0.10$ keV~\cite{2001A&A...365L..67A},
and the peculiar velocity is known, 
$v_p = -29 \pm 299$ km/sec~\cite{2001MNRAS.321..277C}.
We consider 3 cases, {\bf a)} delta function for the spectral band shape,
{\bf b)} Gaussian shape, and {\bf c)} flat (top-hat) shape. 
For the Gaussian and top-hat shapes we use the band widths from Table~1
in~\citeasnoun{Battistelli:2003eh}.
We find
{\bf a)} $log(y)=-4.081^{+ 0.047}_{-0.048}$, 
{\bf b)} $log(y)=-4.080^{+ 0.039}_{-0.055}$, and
{\bf c)} $log(y)=-4.077^{+ 0.040}_{-0.054}$. We thus see, that the 
differences between the filters have almost no effect on
the central value of the Comptonization parameter, and fairly small
effect on the error-bars of $log(y)$.
This is good news (and in agreement with the findings 
of~\citeasnoun{2003ApJ...582L..63C}), 
because the use of filters makes the computation
much longer since one effectively must calculate at many more 
frequencies.
Leaving $v_p$ as a free parameter to be maximized over
has little effect, but leaving
$T_e$ as a completely free parameter instead will increase the error-bars
on $log(y)$ by almost $50\%$.

Using Gaussian filters and assuming the peculiar velocity is measured,
$v_p = -29 \pm 299$ km/sec~\cite{2001MNRAS.321..277C}, one finds that COMA
has a temperature of $T_e = 6.6 \pm 13$ keV, as seen on figure~1a.
This is in good agreement with the X-ray determination, but still
with much larger error-bars. In comparison, the first cluster temperature
measurement using purely the SZ observations gave $T_e = 26 ^{+34}_{-19}$ 
keV for A2163~\cite{2002ApJ...573L..69H}, where only observations at
4 frequencies were available.
Similarly, using the observed temperature, 
$T_e = 8.25 \pm 0.10$ keV~\cite{2001A&A...365L..67A},
one finds the peculiar velocity
$v_p = -300 \pm 500$ km/sec, as seen on figure~1b, which is in good
agreement with the findings of ~\citeasnoun{2001MNRAS.321..277C}.

%\begin{figure}[tbh]
%\begin{center}
%\includegraphics[width=4.5cm,angle=0]{fpecvel.ps}
%\end{center}
%\caption{Using \sasz to determine the peculiar velocity for COMA.
%The figure shows $\Delta \chi^2$ as a function of the peculiar velocity, $v_p$.
%Other parameters have been maximized over. The $1\sigma$ error-bar on
%$v_p$ (corresponding to $\Delta \chi^2 = 1$) gives $v_p = -300 \pm 500$
%km/sec.}
%\label{fig:coma.pev.vel}
%\end{figure}

\section{Conclusion}
We are presenting a user-friendly tool for parameter extraction from
Sunyaev-Zeldovich effect observations. The tool \sasz is based on 
the stochastic method of simulated
annealing, and is useful for any number of observing frequencies and
any spectral band shape. The first version of the tool allows a
determination of the Comptonization parameter, $y$, the peculiar
velocity, $v_p$, and the electron temperature, $T_e$.
The f77 code \sasz can
be readily downloaded together with a user guide with 
examples from\\
{\tt  http://krone.physik.unizh.ch/\~{}hansen/sz/} .

%\del{
\section*{Acknowledgements}
It is a pleasure to thank Sergio Pastor and Dima Semikoz
for collaboration on the fitting formulae, and Nabila Aghanim
for discussions and pointing me towards the data from COMA.

%}

%\appendix

\end{document}